# Tractor beam on the water surface

Horst Punzmann, Nicolas Francois, Hua Xia, Gregory Falkovich[1], and Michael Shats

*Research School of Physics and Engineering, The Australian National University, Canberra, ACT 0200, Australia*

[1] *Weizmann Institute of Science, Rehovot 76100 Israel and Institute for Information Transmission Problems, Moscow 127994 Russia*

**Can one send a wave to bring an object from a distance? The general idea is inspired by the recent success in moving micro particles using light and the development of a tractor beam concept[1,2,3]. For fluid surfaces, however, the only known paradigm is the Stokes drift model, where linear planar waves push particles in the direction of the wave propagation[4,5,6]. Here we show how to fetch a macroscopic floater from a large distance by sending a surface wave towards it. We develop a new method of remote manipulation of floaters by forming inward and outward surface jets, stationary vortices, and other complex surface flows using nonlinear waves generated by a vertically oscillating plunger. The flows can be engineered by changing the geometry and the power of a wave maker, and the flow dissipation. The new method is robust and works both for long gravity and for short capillary waves. We use a novel method of visualizing 3D particle trajectories on the surface. This letter introduces a new conceptual framework for understanding wave-driven flows. The results form the basis for remote manipulation of objects on the fluid surfaces and will help better understanding motion of floaters on the ocean surface and the generation of wave-driven jets.**

What is perceived as fluid motion on the surface perturbed by waves is a motion of the surface shape, not the fluid flow along the surface[7]. Trajectories of fluid parcels on the surface were described analytically only for progressing small-amplitude planar waves, where particles move in the direction of wave propagation[1,3,5,8] along the prolate trochoid (a curve with loops). However, planar linear waves are rare in nature and in laboratory since finite-amplitude waves on the surface of deep water are unstable with respect to amplitude modulation, phenomenon known as the Benjamin-Feir instability[9]. It has been realized theoretically[10,11,12] and confirmed experimentally[13] that 2D waves of finite amplitude develop into 3D waves forming complex wave patterns. What remains unknown is how particles move on the surface of such wave fields. Recent progress in flow visualization and ability to track Lagrangian trajectories in laboratory experiments opens new opportunities to address the problem of the wave-driven surface flows using these tools. In our experiments we generate progressing waves using vertically moving plungers which are periodically inserted in the water. The wave fields are visualized using diffusive light imaging technique[14] and a fast video camera. 3D fluid particle trajectories are tracked using a novel method developed by us and described in the Supplementary Information.

If one uses an elongated wave maker, propagating waves have oval wave fronts with long nearly planar parts, as seen in Fig. 1a for a cylinder (a closer look however reveals that the wave fronts are modulated even at this relatively low amplitude). The maximum of the wave amplitude is at the center of the cylinder side. Here, floating particles are pushed in the direction of the wave propagation forming strong outward jet as seen in Fig. 1b. A compensating return flow is converged towards the sides of the wavemaker.

The flow changes dramatically when the wave amplitude is increased above the modulation instability threshold[15,16], at only 20-30% higher acceleration of the wave maker. As the modulation grows and the cross-wave instability breaks the wave front into the trains of propagating wave pulses, the wave field becomes 3D, Fig. 1c (see also the SI video "Cross-wave development.avi"). Simultaneously, the direction of the central jet reverses. It now pushes floaters inward, towards the wave maker and against the wave propagation! The flow is strong enough to move small objects on the water surface, for example a ping-pong ball, as demonstrated in Figs. 1e,f. The motion of the floater can thus be reversed by simply changing the amplitude of the wave maker oscillations.

The effect of the reversal of the central jet from the cylindrical wave maker is independent of the wave frequency. We find that the tractor beam is produced by the nonlinear 3D waves in the range from long gravity waves (8 Hz) to short capillary waves (50 Hz). The large-scale structure and the topology of the flow are qualitatively the same in both cases, weakly and strongly modulated waves: a quadruple pattern made of 4 large counter-rotating vortices, Figs. 2c-d. The direction of the vortex rotation however reverses with the increase of modulation. This reversal at higher wave amplitudes is always correlated with the generation of stochastic Lagrangian trajectories within a flow region in front of the wave maker, Fig. 2e. This complex chaotic flow efficiently transports fluid in the direction perpendicular to the propagation of the wave pulses. This stochastic pumping seems to be responsible for the evacuation of surface fluid parcels sideways from the wave maker. This transport is compensated by the fluid flow in the inward directed jet.

To understand how stochastic region forms in front of the cylindrical plunger we perform experiments using various shapes of the plungers. If one uses a conical wave maker at low amplitudes, linear waves with circular wave fronts are produced, Fig. 3a. Above the instability threshold, at higher acceleration, the wave amplitude becomes modulated in the transverse direction forming complex 3D wave field, similarly to the cylinder case, Fig. 3b. The wave field is made of periodic pulses of varying phases propagating away from the cone in the radial direction. The surface flow pattern in this case is rather complex, exhibiting stationary vortices of different sizes driven by the propagating wave pulses, Fig. 3c. Strongest small-scale vortices are sustained in the vicinity of the wave maker. They interact with each other forming second and third chains of larger vortices further away from the cone.

These larger vortices can be suppressed by increasing the liquid viscosity. This allows better visualization of the underlying mechanism of vorticity generation by propagating waves. By adding sucrose (the viscosity of a 45% solution is 9.3 times that of water) we reduce significantly the number of vortices in the surface flow. The visualization of the 3D floater trajectories reveals that particles which are close to the maxima of the wave pulses move away from the wave maker until they drift sideways and then reverse to move towards the wave source along a trajectory where wave heights are lower, as seen in Figs. 4a-c. The shapes of the 3D trajectories are not trochoids, expected from the linear wave models, but rather are complex modulated trajectories. Furthermore, close to the wave maker particles move a distance comparable to the wavelength during a wave period, indicating no resemblance to the classical Stokes drift picture.

Strong outward jets are thus formed along the wave maxima. These jets are clearly seen in the averaged particle streaks photographs and in the velocity fields, Fig. 4d-e. The jets diverge and form stationary vortices. The vorticity generation here is somewhat similar to that by the Kelvin-Helmholtz instability in the Rayleigh-Taylor turbulence[17]. This divergence of jets can be visualized by computing the Lyapunov exponents of the Lagrangian trajectories[18], Fig. 4f. The ridges of the maxima of the Lyapunov exponents mark the boundaries of the jet stability. Lagrangian trajectories within outward jets diverge. Since such trajectories are azimuthally periodic, adjacent diverging jets form stable closed vortices.

The wave-jet-driven vortices always form in the near field of any wave maker as soon as waves there become unstable. In the case of the conical plunger the longer axes of these elliptical vortices point in the radial direction, while in the case of a cylinder, ellipses are parallel, so that they interact stronger. The Lagrangian chaos in the near field of the cylinder appears a result of the interaction between wave-jet-driven vortices.

Thus the main ingredients leading to the formation of the tractor beam (inward jet) in the nonlinear wave regime are as follows: (1) modulation instability of the finite-amplitude waves and the onset of the cross-wave, (2) the generation of the spatially periodic outward jets in the near field of the wave maker which sustain stationary counter-rotating vortices, (3) the interaction between jet-driven vortices and the onset of a region of stochastic Lagrangian transport in front of the wave maker, (4) stochastic pumping of fluid parcels out of the turbulent region, and (5) the generation of the large-scale vortex pattern with an inward return flow.

The above results put in question the usefulness of the Stokes drift paradigm for description or prediction of the wave-generated flows in the capillary-gravity range of frequencies. Indeed, in all examples of the wave-generated flows considered here, there are extensive regions where particle motion contradicts not only the value of the Stokes drift velocity, but also its direction. During the initial stage of the flow development, described in the SI, particles at the cylinder side move in the direction of wave propagation. However even during this very short time their velocities differ substantially from the Stokes drift velocity and do not follow a $V \sim a^2$ trend (where $a$ is the amplitude of the damped propagating wave) expected for the Stokes drift. After about 10 wave periods flows start self-organizing into steady-state patterns.

The key for designing controlled jets on the water surface is to break the symmetry of the flow near a wave maker. In the above example the symmetry is broken by the wave three-dimensionality due to the instability. But it can also be broken in a prescribed way by changing the geometry of a plunger. Two-fold symmetry of the cylindrical plunger discussed above, Figs. 2a-b, is just one example. We also produce robust three-fold and four-fold patterns using triangular and square pyramids. All these methods work both for long gravity and short capillary waves and they are versatile enough to manipulate objects on the water surface.

**Acknowledgements**

This work was supported by the Australian Research Council's Discovery Projects funding scheme (DP110101525). HX would like to acknowledge the support by the Australian Research Council's Discovery Early Career Research Award (DE120100364). The authors thank Kamyl Szewc for developing the code for the finite-time Lyapunov exponent analysis used to generate Fig. 3c, and Mark Gwynneth for his help with experimental setup. N. F. acknowledges help of S. Ramsden of the National Computational Infrastructure, Vizlab, ANU with visualization of 3D flows and trajectories using Houdini animation application software. GF research was supported by ISF, BSF and the Russian Ministry of Education.

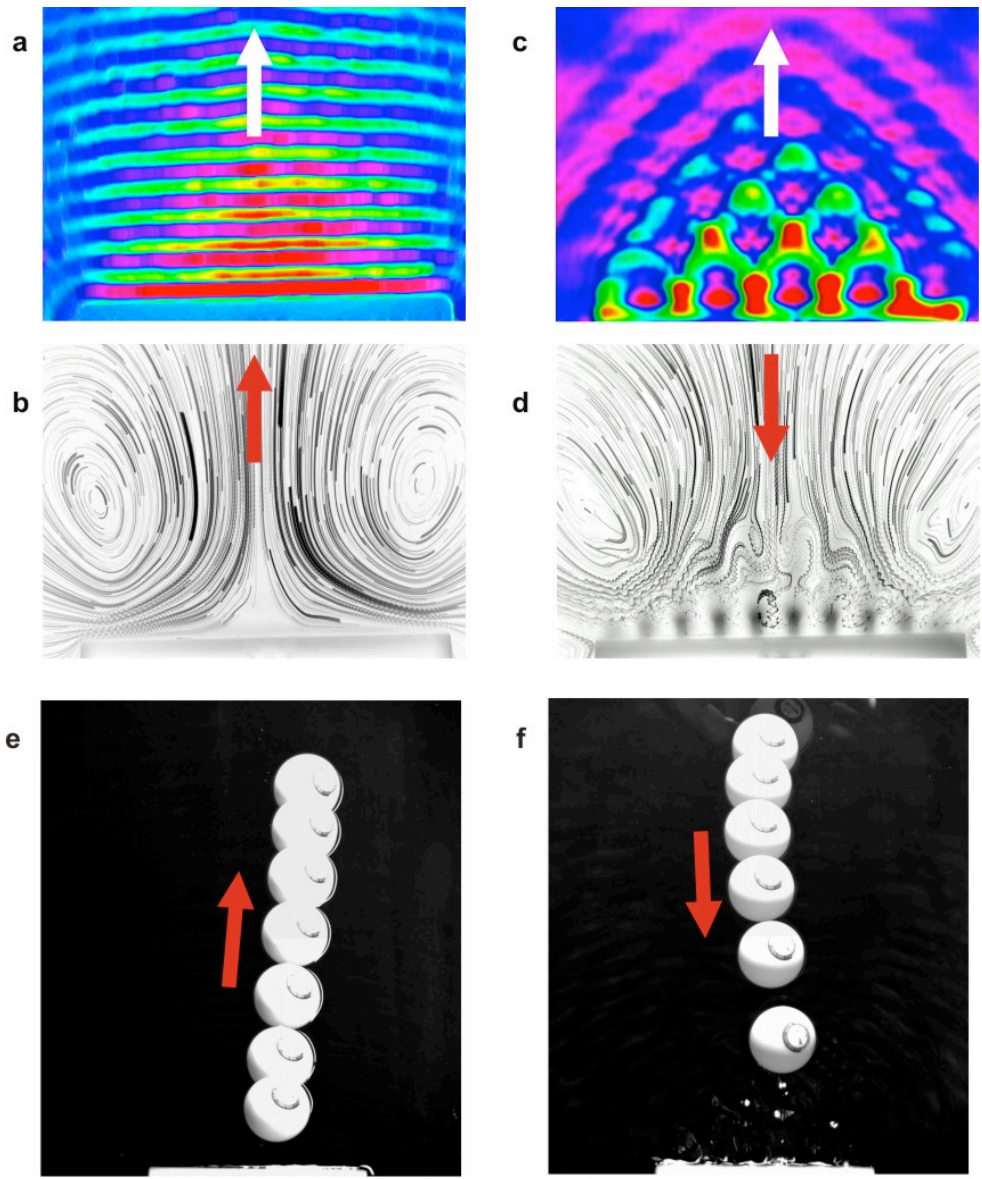

**Figure 1 | Wave fields and surface flows produced by the cylindrical wave maker.**
White arrows show the direction of wave propagation, red arrows show the direction of the surface flow and the ball motion.
**a,** Wave field produced by the cylindrical wave maker at lower acceleration shows nearly planar waves propagating away from the wave maker (seen at the bottom of the panel). **b,** These waves produce strong outward jet in the direction of the wave propagation and the return flows towards the edges of the cylinder. **c,** As the wave maker acceleration is increased (by 30% in comparison with **a**) the modulation instability destroys the wave planarity generating 3D wave field shown here. **d,** This 3D wave field generates strong *inward* directed jet towards the center of the cylinder and the outward return flows away from the edges of the cylinder. **e,** Consecutive positions of the ping-pong ball (2.7 g, 40 mm diameter) placed on the water surface perturbed by a linear wave (as in **a**). Time interval between frames is 0.87 s. **f,** Motion of the ping-pong ball in the inward jet (**c,d**). Time interval between frames is 3.3 s.

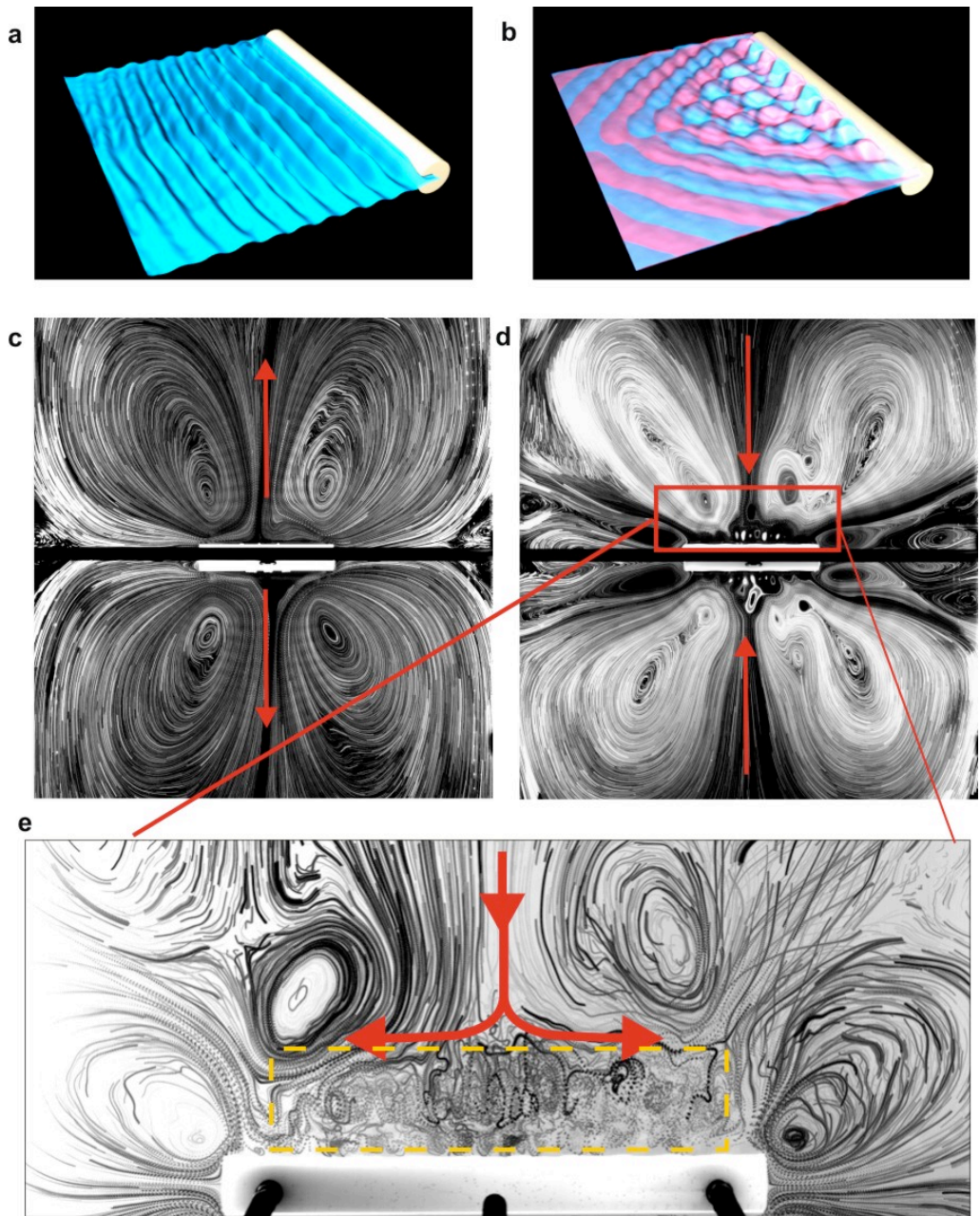

**Figure 2 | Topology of waves and flow patterns produced by the cylindrical wave maker.** 3D reconstruction of waves propagating away from a cylindrical plunger **a,** below, and **b,** above the threshold of modulation instability (Pink and blue wave fields correspond to two consecutive periods of the wave maker phase). **c,** A quadruple pattern is formed around the cylinder made of 4 counter-rotating vortices. Two jets develop in the direction away from the wave maker. **d,** As the wave maker acceleration is increased by 20%, the quadruple pattern is still observed, however the direction of the central jets reverses, towards the wave maker. **e,** Particle streaks in the vicinity of the cylindrical wave maker visualize a region of Lagrangian stochastic transport (yellow dashed box). Turbulence pumps particles away in the direction of red arrows, orthogonal to the wave propagation.

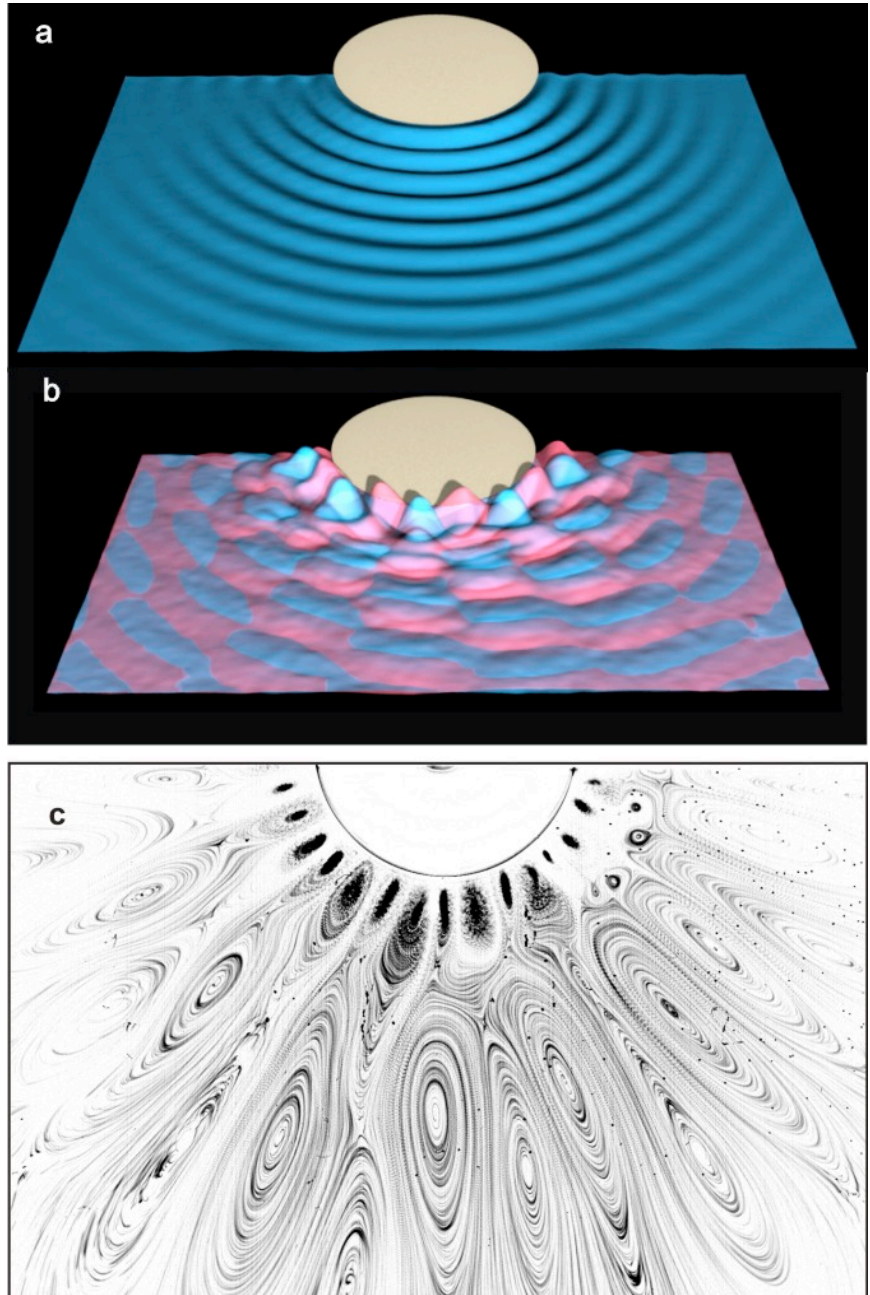

**Figure 3 | Surface flow patterns and wave fields produced by the conical wave maker.**
**a,** Propagating circular waves are produced by a symmetric conical plunger and visualized using diffusive light imaging[14]. **b,** At higher acceleration of the wave maker circular outgoing wave bifurcates into periodic radially propagating pulses as a result of the modulation instability[16]. Pink and blue wave fields correspond to two consecutive periods of the wave maker phase. **c,** Surface flow pattern produced by the 3D wave field shown in **b**. A chain of the spatially and temporally stationary vortices forms near the wave maker. Larger vortices are generated further away from the cone.

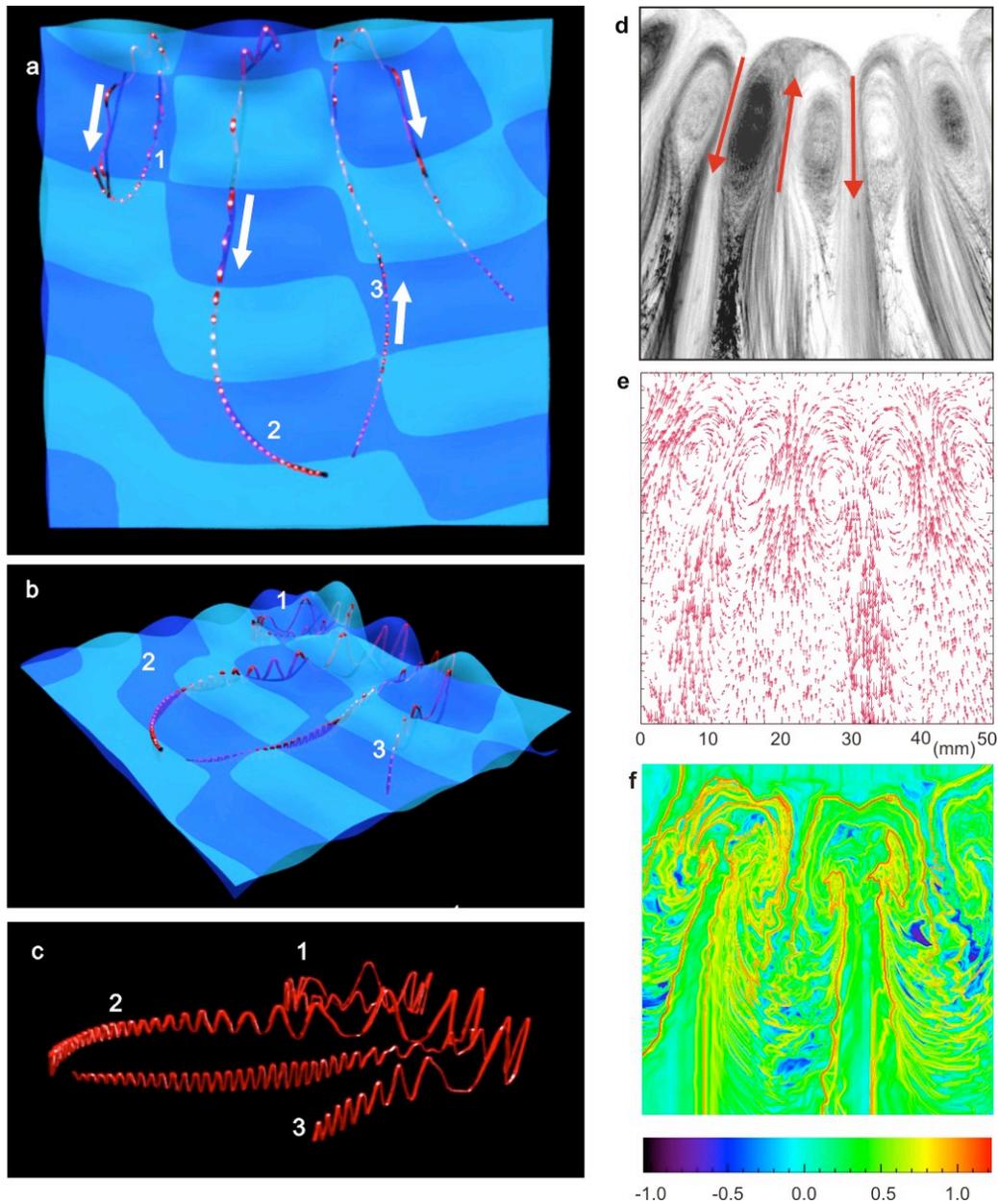

**Figure 4 | Flow structure and floater trajectories in the near field of the conical wave maker.**
**a,** Top view of the 3 particle trajectories. The wave field corresponds to that in Figs. 3b. Trajectory #1 illustrates a particle motion within a smaller vortex. Trajectory #2 shows an outward propagating particle, while the trajectory #3 shows a particle travelling towards the wave maker along the wave minima and then turning radially outward along the wave maxima. **b** and **c**, are the same trajectories shown from different angles. **d,** Time averaged particle trajectories near the wave maker (a cone is on top) show azimuthally periodic strong outward radial jets (red arrows). Between these jets are counter-rotating stationary vortices. **e,** Velocity field of this flow measured using particle image velocimetry method. **f,** Finite-time Lyapunov exponent field visualizes regions of the largest divergence of adjacent Lagrangian trajectories in the flow. Red lines mark the ridges of the maximum of the Lyapunov exponent. These lines outline outward jets seen in **d** and **e**.